\documentclass[12pt]{iopart}
\usepackage{epsfig}

\newcommand{\pbar}      {\mbox{$\overline{p}$}}
\newcommand{\Hbar}      {\mbox{$\overline{H}$}}

\begin{document}

\title[Dense Antihydrogen for Propulsion]{Dense Antihydrogen: Its
Production and Storage to Envision Antimatter Propulsion}

\author{Michael Martin Nieto\dag\, Michael H. Holzscheiter\ddag\, and Thomas
J. Phillips*} 

\address{\dag\ Theoretical Division MS-B285 and \ddag\ Physics Division 
MS-H803, Los Alamos National Laboratory, University of California,
Los Alamos, NM 87545 USA}

\address{* Physics Department, Duke University, Durham, NC 27708
USA}
\vspace{2mm}
\address{Email: mmn@lanl.gov, mhh@lanl.gov,
phillips@physics.duke.edu}


\begin{abstract}
We discuss the  possibility that dense antihydrogen could provide a
path towards a mechanism for a deep space propulsion  system.   We
concentrate at first, as an example, on Bose-Einstein Condensate 
(BEC) antihydrogen. In a Bose-Einstein Condensate,
matter (or antimatter) is in a coherent state analogous to photons in
a laser beam, and individual atoms lose 
their independent identity.  This allows many atoms to be stored in a
small volume. In the context of recent advances in producing and
controlling BECs, as well as in making 
antihydrogen, this could potentially provide a revolutionary path
towards the efficient  storage of large quantities of antimatter,
perhaps eventually as a cluster or solid. 
\end{abstract}





\section{Introduction}

Within this century, the challenge to develop deep space missions will
be undertaken.  To complete a mission within a reasonable time frame,
even to the nearest objects of interest, 
the Oort Cloud or the Alpha Centauri star system (4.3 light years
away), the velocity of the spacecraft needs to be high, up to more
than 10\% of the speed of light.  To achieve 
this one needs the highest energy-density fuel conceivable.  This
would be antimatter; a large amount of it and in a compact form. 

Antimatter can produce three orders of magnitude more energy per gram
than fission or fusion and ten orders of magnitude more energy than
the chemical reactions currently used for 
propulsion. As a result, it is a prime candidate for use in future
exploration beyond the solar system. It also is a candidate for future
missions to the edge of the solar system, 
which now require on the order of 15 years after launch to reach Pluto.

An ideal antimatter propulsion system would convert all the energy
from matter-antimatter annihilation into propulsion. Of course, it has
long been realized that only a fraction of 
the total energy could be used in an antimatter propulsion system, but
much thought has also gone into optimizing this fraction
\cite{rand}. Further,  such a system would require 
what is by today's standards a very large amount of stored
antimatter. For a 500 kg probe, it would take more than 5 g of
antimatter to get out to 200 Astronomical Units (past the 
Heliopause) in 1 year and more than 2.5 kg of antimatter to go 1 light
year in 10 years. 

Alternative propulsion schemes have also been proposed that use
antimatter as a catalyst.  They would  require substantially less
antimatter.  (For example,  they would use 
antimatter to drive fusion reactions \cite{MICF} or fission reactions
\cite{ACMF}.) But for them the total weight of nuclear fuel for the
same energy output is again roughly 3 
orders of magnitude larger than for a pure annihilation energy source.

The amount of antimatter that can be stored per unit volume is
critical to make the necessary equipment of reasonably low mass and to
minimize the requirements on secondary 
technical support systems; i.e., cryogenic
refrigerators. Unfortunately, the density of the most obvious
candidate for a significant amount of stored antimatter, antiprotons,
is limited by the Brillouin density \cite{brillouin},
\begin{equation}
n_0 = \frac{B^2}{8  \pi m c^2}.
\end{equation}
This is the limiting density  for radial confinement in a magnetic
field.  For antiprotons stored in a 6 T magnetic field, typical for
today's technology, it would be around $2 \times 10^{12}$ cm$^{-3}$. 
Therefore, to achieve high storage densities, it will be necessary to
make and store neutral antimatter. The only realistic neutral
candidate is antihydrogen.

Further, a Bose-Einstein Condensate (BEC) 
of antihydrogen, even with present day BEC hydrogen
densities, would be orders of magnitude more dense than the Brillouin
storage density limit for charged 
antiprotons. This latter is itself orders of magnitude higher than the
highest antiproton density so far achieved, $\sim 10^6$ cm$^{-3}$
\cite{forever,ps200}. To make a BEC of 
antihydrogen would be an individual step, where one could learn the
techniques of controlling a relatively large amount of antihydrogen.
The clear ultimate goal would be to make 
very dense antihydrogen,  in the form of clusters or solids (perhaps
stored diamagnetically). 

Therefore, since a) controlled antiprotons have been created, b) cold
antihydrogen has just been made, c) controlled antihydrogen is
envisioned in the foreseeable future, and  d) 
high-density  BEC hydrogen has also been
made (see the next paragraph), we find it appropriate to consider the
possibility of producing antihydrogen BEC as 
a first, but critical, step towards the potential use of antimatter in
a propulsion system. 

Fundamental quantum theory predicts that there are ordinarily two
kinds of particles; i) Fermions, which are particles with half-integer
internal quantum angular momenta (like 
electrons and protons) and ii) Bosons, which are particles with
integer internal quantum angular momenta (like photons or even numbers
of Fermions such as the hydrogen atom).  From 
the early work of Bose and Einstein, it was realized that many Bosons,
if cool enough, could all fall into the lowest allowed quantum energy
and behave as one single (even 
macroscopic) quantum object with remarkable properties.  Two
well-known examples are superconductivity, a sea of paired electrons,
and superfluids, usually a liquid of $^4$He atoms 
(two protons, two neutrons, and two electrons).  The recent discovery
that atomic hydrogen can be condensed into a Bose-Einstein state,
opened up new possibilities. 

The purpose of this communication is to present an initial evaluation
of  possible routes towards efficient production and long-term storage
of antimatter and  also to identify the 
key issues that need to be addressed, specifically using BEC
antihydrogen. We note that much of the equipment and the techniques
that would be needed to study some of these concepts 
are already commonly available. Further, because of the symmetry
between matter and antimatter, one should be able to perform many of
the studies with matter.  This is important 
since antiprotons are not (yet) readily available in large quantities.


\section{Antihydrogen}

Antihydrogen is the antimatter equivalent of hydrogen.  It is composed
of a positron in an atomic orbit around an antiproton. Antihydrogen is
produced if a free antiproton 
($\overline p$) picks up a positron ($e^{+}$) in an atomic orbit.

The antiprotons for antihydrogen can be made in high-energy
collisions. They are currently being produced at the Fermilab and CERN
particle accelerators for use in high-energy 
physics experiments. Antiprotons at CERN have been slowed and captured
in Penning traps \cite{large_trap, GG_cool_pbar}. The collection
process is currently quite inefficient, but 
this efficiency could be improved by several orders of magnitude with
straightforward changes \cite{GPJ01}.  The positrons for antihydrogen
are much more easily made using either 
small electron accelerators  \cite{positrons}  or collected from
certain radioactive sources with well-known gas moderation techniques
\cite{trap_positron_89}. 

Previously, a few atoms of antihydrogen have been produced at CERN
\cite{HbarCERN} and Fermilab \cite{HbarFNAL}, but these have been
produced at relativistic speeds, much too fast 
to capture. However, the ATHENA experiment at CERN recently produced a
significant amount (of order 50,000) of cold antihydrogen atoms
\cite{coldhb}, a feat which caused 
international excitement \cite{athenapr}. Later, the ATRAP
collaboration also reported production of antihydrogen \cite{atraphb}. 

The antihydrogen program at CERN  \cite{athena,atrap} is geared
towards  tests of fundamental physics principles such as spectroscopic
studies of CPT invariance and the weak 
equivalence principle. While the aim of that research is to produce,
what is for our purposes, a moderate number of anti-atoms, much can be
learned from it about the different 
production and storage mechanisms proposed for our present purposes.


\section{Production of Cold Antihydrogen}

     For any scheme proposed to produce slow antihydrogen, the velocity of
the positron and antiproton relative to each other must be very
small. This requirement is best described in terms of a temperature of
the order of several Kelvin, describing the 
energy distribution in the center-of-mass frame. Such temperatures can
be achieved by trapping the positrons and antiprotons in Penning
traps.  These traps confine charged 
particles in the radial direction with a strong solenoidal magnetic
field (typically of the order of several Tesla), and in the
longitudinal direction with  static electric fields. 

Once either positrons or  electrons have been captured in a Penning
trap, they cool very rapidly to equilibrium with the ambient
temperature of the trap by emitting synchrotron 
radiation. In a 6 Tesla magnetic field the time constant for this
process is approximately one hundred milliseconds. 

Antiprotons have a much slower time constant for cooling through
synchrotron radiation due to the heavier mass, but they can be cooled
by loading electrons in the same trap. Coulomb 
collisions between the electrons and the antiprotons lead to
thermalization of the two species, thus cooling the antiprotons. The
heat input to the cold electrons due to this 
process is continuously removed by synchrotron radiation into the heat
sink of the cryogenic environment of the experimental
apparatus. Therefore, the velocities of all particles 
involved in this process can be reduced by cooling the physical
structure of the trap to cryogenic temperatures. 

This technique was first introduced by Rolston {\it et al.}
\cite{rolston} and has been successfully used by two experiments at
CERN, PS196 \cite{ps196} and PS200 \cite{ps200}. The 
latter experiment currently holds the world record for the number of
cold antiprotons confined in a trap of 1 million \cite{forever}. The
PS200 technology has now been fully 
incorporated into the ATHENA experiment at CERN \cite{coldhb,athena}.


\section{Processes to Combine Antiprotons and Positrons into Antihydrogen}

When an antiproton and a positron come together, they can only become
an antihydrogen atom if there is a mechanism to remove energy from the
system.  This binding energy can be 
removed either by  the radiation of a photon (radiative recombination)
or it can be carried off by a third particle (three-body
recombination). A number of different channels have 
been considered for such processes \cite{rpp}.

One of the strongest candidates for our purpose appears to be the
three-body recombination (TBR) reaction $\pbar + 2e^+ \rightarrow
\Hbar + e^+$. The calculated rate for this 
reaction can be quite high \cite{3bodyBfield}:
\begin{equation}
\Gamma = 6 \times 10^{-13} \left({4.2\over T}\right)^{9/2} n_e^2\: 
     [{\mathrm{s}}^{-1}]  \label{rate1}
\end{equation}
per antiproton, where $n_e$ is the positron density in number per
cm$^3$ and $T$ is the temperature.  For a positron density of $10^{8}$
cm$^{-3}$ at $4.2^\circ$ K the half life for 
an antiproton in the positron plasma before it is converted into a
antihydrogen atom is $170 \mu $s. This essentially promises an
instantaneous conversion of all antiprotons into 
antihydrogen atoms. 

But most of these atoms are formed in highly
excited Rydberg states, and it currently is not clear what will be 
the effect of the electric and magnetic fields in the 
trap environment on the net production rate. The recombination
rate observed to date by the ATRAP collaboration is significantly
lower than the predictions from equation (\ref{rate1}). 
Most likely this is due to field ionization of the higher lying Rydberg
states. To overcome this problem and to increase the net production
rate it will be necessary to de-excite these 
atoms with an appropriate laser. Care must then be taken that
photo-ionization of the Rydberg atoms does not limit the net
production. 

A second production mechanism of interest to us is the spontaneous
radiative recombination (SRR) in which a positron is bound to an
antiproton through emission of a photon. In this process 60\% 
of the antihydrogen atoms are formed in states $n =1$ through 
$10$ and 20\% of the atoms are formed directly in the ground state. The
recombination coefficient 
$\alpha^{SRR}  = \langle \sigma^{SRR}(v) v \rangle$ 
for this process (at a positron
temperature of 0.1 meV) is of the order of $10^{-10}$
cm$^3$/s.  This recombination process is therefore 
substantially slower than the theoretical three-body reaction at the
same low temperature. But it produces atoms stable against
reionization by the strong electromagnetic fields in 
the environment. Additionally, the recombination rate can be enhanced
by using a laser to stimulate a transition into a specific, stable, 
low-lying $n$ level. 

Assuming the parameters proposed for the ATHENA experiment of $10^7$
antiprotons and $10^8$ positrons confined to a volume of 1 cm$^3$,
this reaction yields a production rate of 90,000 antihydrogen 
atoms per second. The entire sample of $10^7$ antiprotons 
would be converted to antihydrogen in less than 20 minutes. 
Considering the 
fact that the long range goal will be to store the produced
antihydrogen for extended periods of time this rate is fully
sufficient for initial experiments. 

Several ideas to enhance the net production rate for both of the above
mechanisms have been discussed in the literature \cite{rpp}. 


\section{Experimental Methods for Antihydrogen Production}

\begin{figure}[ht]
    \centering
    \noindent\mbox{\epsfig{file=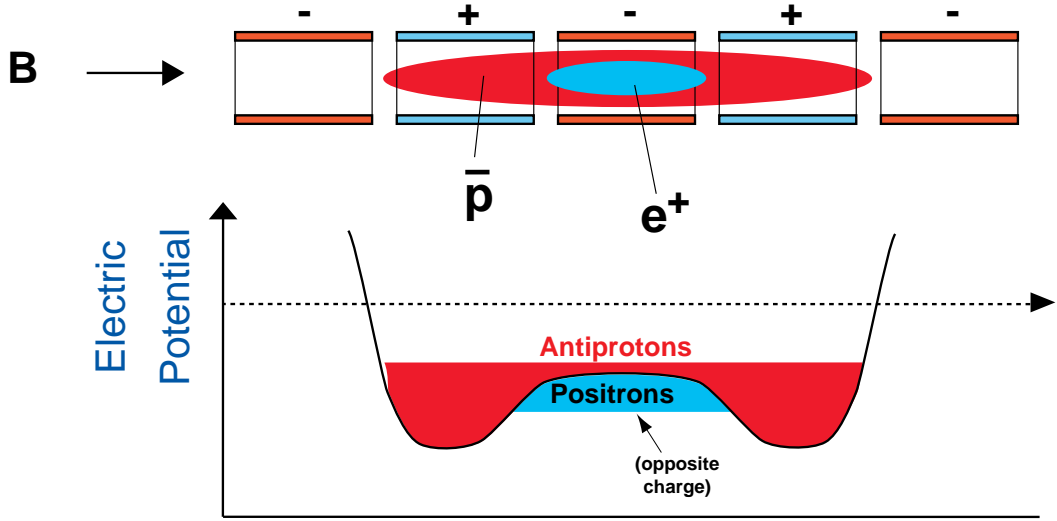,width=5.5 in}}
   \caption{Physical realization and the electric potential of a nested
    trap to store antiprotons and positrons in the same location.
    \label{fig:nestedtrap}}
\end{figure}

Positrons and antiprotons have opposite charge, so they must be
trapped in potential wells of opposite signs in Penning traps.
Several ideas have been discussed in the literature 
on how to make the two oppositely charged clouds physically overlap in
space, either in static configurations or by moving the electric wells
through one another. The simplest rendition of such a scheme is the
``nested trap" \cite{nested_trap}. In its most straight forward
realization it consists of a 
sequence of five cylindrical electrodes that are biased in such a way that 
an elongated potential well is formed for antiprotons. 
This well which has a hill
in its center that can then serve as a well for the oppositely
charged positrons.  (See Figure \ref{fig:nestedtrap}.)


To overcome the possible separation of the oppositely charged plasmas,
the ATHENA collaboration uses an approach where the antiproton cloud
is released from a side well into a 
positron plasma.  The density and length of the positron plasma is
chosen so that an individual low-energy antiproton entering the
positron plasma along the axis will come to rest 
before leaving the plasma column again.  Therefore, it will recombine
with high probability in a single passage. 

\begin{figure}[ht]
    \centering
    \noindent\mbox{\epsfig{file=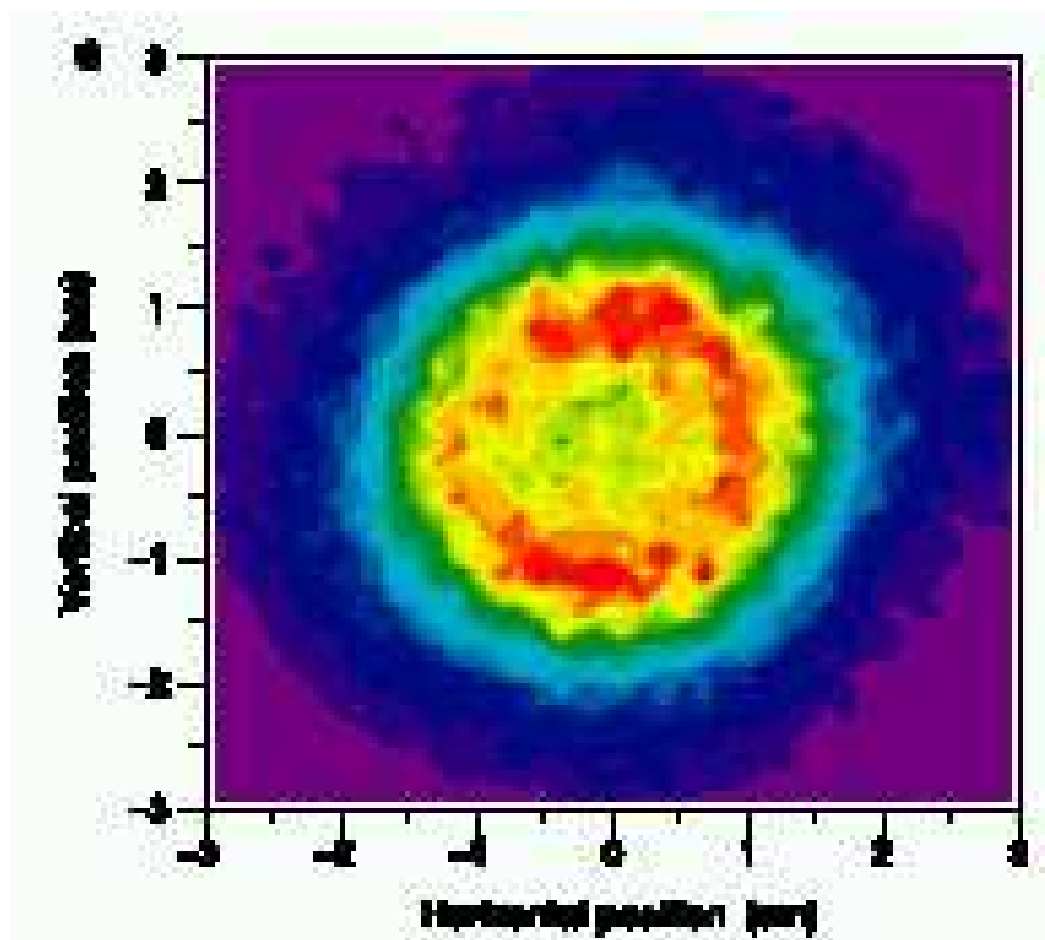,width=5.5 in}}
   \caption{Contour plot of the distribution (obtained by projecting
    onto the plane perpendicular to the magnetic field) of the vertex
    positions of reconstructed anti-hydrogen annihilation events from
    the ATHENA cold antihydrogen production experiment \cite{coldhb}.
    \label{antihdie}}
\end{figure}

All of the above considerations have been successfully incorporated in
the ATHENA experiment, which recently detected the creation of around
50,000 cold anti-hydrogen atoms at CERN 
\cite{coldhb}. The creation was demonstrated by the detection of the
annihilation products, preferentially on the walls of the trap.  (See
Figure \ref{antihdie}.)  This completes a major 
step on the road to the envisioned technology.


Once antihydrogen has been formed it will
no longer be trapped in the Penning trap configuration.  It will
escape in the direction of the antiproton's initial velocity and most
likely will annihilate at the wall of the 
trap. This will make it difficult to diagnose the dynamics of the
recombination reaction and the dependence of the recombination rate on
the many different parameters relevant in 
these experiments. While recombination at rest will be an important
part of an experiment aimed at filling some form of neutral storage
scheme, we feel a method allowing easier 
diagnostics schemes is appropriate for the first stages of development.

Observe that it is the relative velocity between positrons and
antiprotons which drives the reaction rate. The velocity of a positron
at 4 Kelvin is about $1.1 \times 10^4$ m/sec. 
An antiproton of the same velocity has an equivalent energy of about
0.6 eV (or 7300 K). This allows one to accelerate antiprotons into or
through positron clouds without 
diminishing the conversion rate to antihydrogen.

\begin{figure}[ht]
    \centering
    \noindent\mbox{\epsfig{file=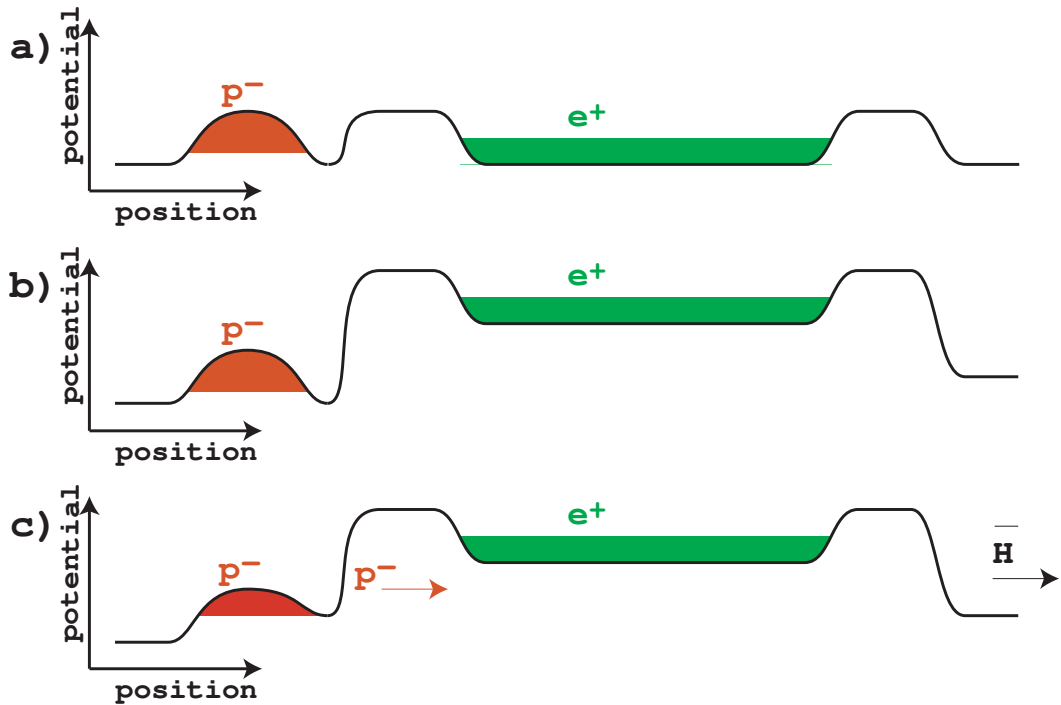,width=5.5 in}}
   \caption{To make a beam of antihydrogen, (a) cold antiprotons and
    positrons are first stored in separate potential wells in a penning
    trap. The figure plots electrical potential on the vertical axis
    but because of their negative charge, the potential energy for
    antiprotons is inverted relative to the electric potential and they
    can be trapped under peaks in the electric potential.  One can also
    look at the figure upside down.  In (b), a small
    voltage is applied between the antiprotons and the positrons.  So, when
    in (c) the right side
    of the potential barrier confining the antiprotons is lowered (raised
    with respect to the positrons),
    the antiprotons are accelerated. They pass through the positron
    plasma with their momentum directed down the axis of the trap.  If
    the density of positrons is high enough, some antiprotons will pick
    up positrons and exit the trap as a beam of antihydrogen.
    \label{fig:MakeBeam}}
\end{figure}

Because of this, an appropriate proposal to make antihydrogen starts
by keeping the antiprotons and positrons in separate potential wells
in a Penning trap, as shown in Figure~\ref{fig:MakeBeam}.
The potential of the antiprotons can be raised with a small
applied voltage so that when the antiprotons are released, 
they are accelerated along the axis of the trap. 
When they enter the positron plasma, some antiprotons will pick up
positrons to form antihydrogen.  The neutral antihydrogen will no
longer be confined, so it will exit the trap 
along the axis, making a beam of antihydrogen.  


For sufficiently large
positron densities, the production efficiency should be very high.  To
produce a well defined beam, the 
initial velocity of the antiprotons and the density and length of the
positron plasma must be carefully chosen.


\section{Antihydrogen Storage}

Long term storage of substantial amounts of antimatter must be
developed to enable space missions relying on antimatter-based
propulsion systems.   Although it is  clear that 
ultimately neutral antimatter must be used,  up to now, no valid
long-term storage concept for large quantities of antihydrogen has
been developed. On the other hand, BEC 
confinement of neutral spin-polarized hydrogen atoms at densities up to
$5 \times 10^{15}$ cm$^{-3}$ has been demonstrated. \cite{BEC}.  We
believe that developing the tools for 
making antihydrogen BEC could be an important step towards solving the
antimatter storage problem. 

While storage of high density samples of neutral alkali atoms has been
demonstrated using magneto-optical and optical traps, the lack of
appropriate laser systems has so far 
precluded hydrogen from such methods. Nevertheless, the BECs with the
largest total number of particles have been formed using
hydrogen. Until now neutral hydrogen has been stored 
either in cryogenic cells, by using wall collisions as the confinement
mechanism, or in magnetic traps employing static magnetic gradient
fields \cite{H-trap-1,H-trap-2}.  Although 
total quantum reflection at containment walls is a possibility, without such a new technology, wall collisions are clearly unacceptable for the case of antihydrogen. One then needs
to consider magnetic traps.

A static magnetic field trap is constructed with a minimum at its
center.  Then the only states that can be trapped are the so-called
low-field seeking states.  (The upper two 
states in the Hyperfine diagram for the ground state of atomic
hydrogen.) These states will decay by the mechanism of dipole
relaxation to the lower, un-trapped states with a rate 
constant  of $\sim 10^{-15} cm^{-3}s^{-1}$ \cite{stoof,evapcool}.
Another (smaller) loss mechanism is three-body recombination, which
for hydrogen is ($\sim 10^{-19}cm^{-3}s^{-1}$) 
\cite{3bodyHdecay}. Apparently, these effects would strongly limit the
density of achievable BECs. 

But there are a number of other considerations that mean a controlled
dense BEC is not ruled  out in principle.  Firstly, it is interesting
to note that in a BEC, the coherence 
properties reduce the above dipole and three-body transition rates  by
factors of 2 and 6, respectively \cite{3bodytheory}.  Such a factor
has been verified experimentally using Rubidium \cite{3bodyexpt}.

Next, Lovelace {\it et al.} showed \cite{lovelace1,lovelace2} that a
trap confining all hyperfine states of neutral (anti)hydrogen can be
constructed using AC magnetic fields.  (The 
situation is similar to strong focusing in particle accelerators.)
Such a trap has not yet been experimentally realized because i)
magnetostatic traps have been entirely 
sufficient for current research interests and ii) the technical
challenges presented in generating AC magnetic fields with the
appropriate amplitude and frequency are sizable. 
However, especially given the proper interest, traps of this type are
certainly feasible. 

But most importantly, the above transition rates are for current types
of traps, with magnetic fields on the order of up to a few T.  The
rates are strongly magnetic-field 
dependent, and disappear for very large magnetic fields, of order 140
T \cite{3bodyHdecay,highB}. Fields on this order can now be made in a
pulsed mode.  Efforts to eventually 
create them statically have commercially reached 21 T \cite{oxford}.

At present the wasteful method of resonant evaporative cooling is used
to achieve the temperatures and densities needed to form a hydrogen
BEC. But the development of lasers for 
direct and efficient cooling of hydrogen atoms has now just
started. Efficient laser cooling of hydrogen will revolutionize the
methodology of forming, controlling, and studying 
hydrogen Bose-Einstein Condensates. All the cooling, as well as the
optical storage, methods that have been so successful with Alkali BECs
could then be used with antihydrogen. That 
is to say, once an antihydrogen atom has been made and stored, in
principle the only additional requirement (beyond those needed for 
normal hydrogen) is the requirement that the residual 
pressure and temperature be low enough so that no residual normal atom
be around to annihilate an antihydrogen. 

Further, enormous advances have been achieved in the control of Alkali
BECs leading to the demonstration of the coherent control of a BEC and
the demonstration of the concept of 
atom-lasers \cite{ketterle}. Based on these developments a concept of
coherently amplified antimatter waves could be envisioned to increase
antihydrogen storage capabilities by a 
significant factor. Such schemes are far beyond current technological
capabilities, but the study of the underlying physical principles is a
necessary beginning if one is to hope 
for a breakthrough discovery.

If the envisioned progress comes to fruition, laser cooling could then
be used in an attempt to efficiently make an antihydrogen BEC. 


\section{Conclusion}

An antihydrogen BEC would be an important  step down a path that could
eventually lead to even more dense anti-hydrogen molecules, liquids,
solids, and cluster ions 
\cite{silvera}-\cite{young}.  Indeed, since one might expect the next
stage to be going from controlled ultra-cold (below 50 $\mu$K) BEC
hydrogen atoms to controlled hydrogen 
molecules, it is heartening that there is evidence of a
hydrogen-molecule superfluid with a critical temperature of 0.15 K
\cite{grebenev}.  Since the triple point of hydrogen is at 
13.8 K, a potential path to denser condensed antimatter becomes more
interesting. 

In our opinion a space-certified storage system for neutral antimatter
can not be obtained from a linear extrapolation of heretofore existing
technologies. Rather, it  requires a 
scientific and/or technological breakthrough of the kind that
Bose-Einstein Condensation might provide. While breakthroughs can
never be predicted, they typically will not happen 
without the definition of a strong need and the challenge presented to
the scientific community by a truly ambitious goal. Meanwhile many of
the underlying issues can be addressed 
with both the modest supply of antimatter available at this time at
accelerator centers world wide and with the limited means to store the
particles. The technological and 
scientific knowledge gained in these tests will enable us to lay out a
path into the future of antimatter-based propulsion systems. 

Current antiproton production rates are low. While clever techniques
can enhance these rates by several order of magnitude and quantities
sufficient for advanced concepts can be 
produced given enough economic and political pressure onto the few
available sources, a real breakthrough can only come through continued
interest and research in this area. A good 
analogy is the comparison between a light bulb and a laser. In both
cases light is produced, but in one system through thermal heating of
a material and in the other through 
coherent processes. Antiprotons are currently produced by heating a
metal target with a primary proton beam. This is a direct analogy to
the light bulb --- we are still awaiting the 
invention of a `laser-equivalent' for the production of particles of
antimatter. 

Similarly, the efficient production and storage of antihydrogen (perhaps
as a BEC, cluster, or solid), which are necessary steps towards large
large-scale 
production as a fuel, can be studied with the 
quantities of antihydrogen soon to be available to us.  The new modest
production at CERN \cite{coldhb} should help us in identifying key
problems and in developing technologies in 
preparation for future higher production rates of antimatter.  

CERN is the only choice for the location to make the initial scientific
advances.  There are a number of steps needed to create and understand
the desired form of antihydrogen.  (i) Trap and cool the antihydrogen in
a magnetic trap, developing tunable Lyman-alpha lasers along the way. 
After this first step one could (iiA)  convert the antihydrogen into a
BEC.  Alternatively, one could (iiB) form and control molecular
antihydrogen, and finally (iii) convert the antihydrogen into a
superconducting cluster and/or solid of antihydrogen.  Each of these
advances, with strong funding, might be made in on the order of 5-10
years.

Future, even moderate, sources of antihydrogen 
(say at GSI or Fermilab) would also enable
research on antimatter-based propulsion.  One could also consider the 
use of antimatter to drive fusion reactions \cite{MICF} or fission
reactions \cite{ACMF} and also study crucial issues necessary to
develop a propulsion system based entirely upon antimatter annihilation.

To meet the challenge of interstellar travel we have to overcome
enormous scientific and technological barriers. Antimatter-matter
annihilation is one of  the prime candidates to 
achieve the high specific impulse  i) desired for the challenging
missions of exploring the Heliopause and  visiting the Oort Cloud, and
ii) needed if we plan to attempt a 
rendezvous with the nearest star systems. While no clear pathway to
the necessary technologies exists, possible concepts can be identified
based upon the need for the highest 
possible energy densities.  

The experimental development of such
systems as BECs and atom lasers in the normal
matter world of laboratory-sized research 
equipment can help us to  reach these most ambitious goals.  To
achieve them quickly it is necessary to set ourselves in motion now. 


\section*{Acknowledgements}

We thank Michael Charlton, Lee Collins, Terry Goldman, TJ Johnson,
Peter Milonni, Slava Turyshev, and 
especially Eddy 
Timmermans for their helpful comments and information. We all (MMN,
MHH, and TJP) acknowledge the support of the 
US Department of Energy.  



\end{document}